\documentclass[aps,pra,reprint,amsmath,amssymb,longbibliography]{revtex4-1}
\usepackage{graphicx}
\usepackage{bm}
\usepackage{amsthm}
\usepackage[colorlinks=true,linkcolor=blue,citecolor=blue,urlcolor=blue]{hyperref}
\hypersetup{pdftitle={Every three-qubit pure state is locally unbiased to a fixed GHZ basis}}
\newtheorem{theorem}{Theorem}
\newtheorem{lemma}{Lemma}
\newcommand{\ket}[1]{\lvert #1\rangle}
\newcommand{\bra}[1]{\langle #1\rvert}

\newcommand{\adj}{\operatorname{adj}}
\newcommand{\im}{\operatorname{Im}}
\newcommand{\re}{\operatorname{Re}}

\newcommand{\C}{\mathbb C}
\newcommand{\su}{\mathrm{SU}}
\newcommand{\so}{\mathrm{SO}}
\begin{document}
\title{Every three-qubit pure state admits an equal-modulus expansion in a fixed GHZ basis under local unitaries}
\author{Wei Song}
\email{wsong1@mail.ustc.edu.cn}
\affiliation{School of Physics and Materials Engineering, Hefei Normal University, Hefei 230601, China}
\author{Xiao-Lan Zong}\email{zxl@hfnu.edu.cn}
\affiliation{School of Physics and Materials Engineering, Hefei Normal University,
Hefei 230601, China}
\begin{abstract}
In general, local unitary transformations alone do not necessarily generate a complete orthonormal basis from a pure state. We prove that any three-qubit pure state, after suitable local unitary transformations, can be expanded in the same fixed Greenberger-Horne-Zeilinger basis, and all expansion coefficients have equal modulus. Thus, eight mutually commuting product reflection operators can generate a complete orthonormal basis from an arbitrary three-qubit pure state, thereby proving the three-qubit commuting-reflection encoding conjecture. We also give two applications of this result: first, using three Bell entanglement links, a pre-specified three-qubit pure state can be deterministically distributed through joint projective measurement and local correction; second, using local random unitary operations, a known pure state can be transformed into a maximally mixed state. The mixing protocol uses a shared three-bit classical label and attains the minimum possible label entropy, even when global unitaries are allowed.
\end{abstract}
\maketitle

\section{Introduction}\label{sec:intro}

A complete orthonormal basis determines a rank-one projective measurement. If the basis vectors are locally unitary equivalent to one another, they possess the same entanglement properties. The Bell basis is precisely such an example and is also the foundation of entanglement swapping~\cite{Zukowski1993}. This naturally raises the question: for a given multipartite pure state, can it be extended to a complete orthonormal basis such that all the remaining basis vectors are locally unitary equivalent to that state?

Tanaka, Markham, and Murao discussed this problem when studying the local encoding of classical information~\cite{Tanaka2007}. The specific form of such encodings depends on which classical label information is available to each party. If each party can choose local operations based only on one private classical bit of their own, then the states that allow complete encoding are exactly the locally maximally entangleable states characterized by Kruszynska and Kraus~\cite{Kruszynska2009}. If all parties can choose local operations based on the full classical label, the range of states for which encoding can be implemented is broader. Pimpel, Renner, and Tavakoli proved several results in bipartite systems and studied the corresponding three-qubit problem~\cite{Pimpel2023}.

Pauwels and G"uhne further studied a class of encodings satisfying an additional condition: all product unitary operators commute pairwise, and each single-qubit factor is a Hermitian involution, i.e., a Hermitian operator whose square equals the identity operator~\cite{Pauwels2026}. They proved that the existence of such an encoding is equivalent to the corresponding state being transformable by local unitary transformations into an equal-modulus superposition state in some graph-state basis, and conjectured that any three-qubit pure state admits such a commuting-reflection encoding. They also proved that certain four-qubit states cannot achieve complete local encoding even without the above commuting-reflection restriction.

Eisfeld, Kampermann, and Wyderka raised a closely related question for a specific three-qubit Pauli group. The common eigenbasis of this group can be transformed by local unitary transformations into the Greenberger--Horne--Zeilinger (GHZ) basis~\cite{Eisfeld2026}. They considered states with equal-modulus expansion in this common eigenbasis but did not resolve whether these states can represent the local unitary equivalence classes of all three-qubit pure states.

This paper provides a complete analytical proof: any three-qubit pure state, after suitable local unitary transformations, can be expanded with equal moduli in the standard GHZ basis. This result proves the commuting-reflection encoding conjecture in Ref.~\cite{Pauwels2026} and also answers the three-qubit universality question raised in Ref.~\cite{Eisfeld2026}. Here the GHZ basis is fixed for all input states; what varies with the input state is the required local unitary transformation and the phases of the expansion coefficients.

The key to the proof is an identity among two-body correlation matrices. When these matrices are all invertible, we prove the existence of the required local bases by using, depending on the case, the intermediate value theorem and an odd-map argument on the
three-sphere. Subsequently, using the density of states with invertible correlation matrices and the compactness of the local unitary group, we extend the conclusion to the case of singular correlation matrices.

Using the orthonormal basis thus constructed, we give two applications. First, a pre-specified three-qubit pure state can be distributed through three Bell entanglement links. Second, local random unitary operations can be used to transform a known pure state into a maximally mixed state and to minimize the entropy of the required random label.

\section{Main result and local encoding}\label{sec:main}

Using $X,Y,Z$ to denote the Pauli matrices, and adopting the convention that operators written side by side represent their tensor product, e.g., $ZZI=Z\otimes Z\otimes I$. Let
$\mathcal X=\{000,001,010,011\}$, and let $\bar x$ denote the result obtained by flipping every bit of the bit string $x$. Fix the following GHZ basis:
\begin{equation}
 \ket{G_{x,\pm}}=\frac{\ket{x}\pm\ket{\bar x}}{\sqrt2},
 \qquad x\in\mathcal X.
 \label{eq:ghz}
\end{equation}
If the moduli of all eight expansion coefficients of a normalized state in this basis are equal to $1/\sqrt8$, we say that the state has an equal-modulus expansion in this basis.

\begin{theorem}\label{thm:main}
For any normalized state $\ket\psi\in(\C^2)^{\otimes3}$, there exist single-qubit unitary operators $V_A,V_B,V_C$ such that $V=V_A\otimes V_B\otimes V_C$ satisfies, for all $x\in\mathcal X$ and both signs $\pm$,
\begin{equation}
 \bigl|\bra{G_{x,\pm}}V\ket\psi\bigr|^2=\frac18.
 \label{eq:flat}
\end{equation}
\end{theorem}

This theorem applies simultaneously to product states, biseparable states, W-class states, and GHZ-class states. Local unitary transformations do not change the entanglement properties of these states; therefore, the fact that the moduli of the expansion coefficients are the same does not mean that different input states have the same entanglement. In such an equal-modulus expansion, the differences in the local unitary invariants of different states are reflected in the relative phases of the expansion coefficients.

Below we use Theorem~\ref{thm:main} to construct a local encoding. Consider the Abelian group generated by three Pauli operators
\begin{align}
 \mathcal S&=\langle ZZI,IZZ,XXX\rangle,\notag\\
 S_s&=(ZZI)^{s_1}(IZZ)^{s_2}(XXX)^{s_3},
 \qquad s\in\{0,1\}^3.
 \label{eq:stabilizer}
\end{align}
This group has eight elements, and their common eigenbasis is precisely the GHZ basis in Eq.~\eqref{eq:ghz}. Since all group elements except the identity have zero trace, combining this with Eq.~\eqref{eq:flat} gives
\begin{equation}
 \bra\psi V^\dagger S_sV\ket\psi=\delta_{s,0}.
 \label{eq:expectation}
\end{equation}
Define the product unitary operators and the encoded states generated by them as
\begin{equation}
 U_s=V^\dagger S_sV,\qquad \ket{\psi_s}=U_s\ket\psi,
 \label{eq:encoding}
\end{equation}
then
\begin{equation}
 \langle\psi_s|\psi_t\rangle
 =\bra\psi V^\dagger S_{s\oplus t}V\ket\psi
 =\delta_{s,t}.
 \label{eq:gram}
\end{equation}
Therefore, these eight encoded states are mutually orthonormal, form a complete basis, and are all locally unitary equivalent to the input state $\psi$.

Each $U_s$ is a tensor product of three single-qubit Hermitian involutions; if there is an overall minus sign, it can be absorbed into one of the local factors. Here it is the overall product unitary operators that are required to commute, whereas the local factors used by the parties need not commute with one another. Each party can choose their own local operation according to the full label $s$, which is consistent with the commuting-reflection encoding model in Ref.~\cite{Pauwels2026}. The GHZ stabilizer is local Clifford equivalent to the stabilizer of a connected three-vertex graph state. Theorem~\ref{thm:main} gives a local encoding for every three-qubit pure state using the same GHZ stabilizer. The local change of basis $V$ may depend on the state.

The correspondence between the results of this paper and those of Ref.~\cite{Eisfeld2026} is as follows. The Pauli group considered in that paper is $\mathcal G=\langle ZZZ,XIX,IXX\rangle$. Let $\mathsf H$ denote the single-qubit Hadamard gate; then
\begin{equation}
 \mathsf H^{\otimes3}\mathcal G\mathsf H^{\otimes3}
 =\langle XXX,ZIZ,IZZ\rangle=\mathcal S,
 \label{eq:group-equivalence}
\end{equation}
where the last equality uses $(ZIZ)(IZZ)=ZZI$. This shows that the common eigenbases of the two groups are related by the same fixed local Clifford transformation, and the corresponding families of equal-modulus superposition states can also be converted into each other through this transformation. Theorem~\ref{thm:main} therefore proves the universality of the state family proposed there.

\section{Proof of Theorem~\ref{thm:main}}\label{sec:proof}

\subsection{A phase condition}\label{sec:phase}

Choose three unit Bloch vectors $a,b,c$. For each party $J=A,B,C$, take the normalized $+1$ and $-1$ eigenvectors of the corresponding Pauli observable, in that order, as the two columns to form the unitary matrix $E_J$. Let
\begin{equation}
 E=E_A\otimes E_B\otimes E_C,\qquad \ket w=E^\dagger\ket\psi.
 \label{eq:localbasis}
\end{equation}
Write $w_x=\langle x|w\rangle$, $\langle O\rangle_w=\bra w O\ket w$, and define
\begin{equation}
 Q(w)=w_{000}w_{011}w_{101}w_{110}
       \overline{w_{001}w_{010}w_{100}w_{111}}.
 \label{eq:Q}
\end{equation}
In each product in Eq.~\eqref{eq:Q}, the local labels $0$ and $1$ each occur twice at every site. The eigenvector phases therefore cancel in $Q$. This shows that $Q$ does not depend on the choice of phases of the eigenvectors, but only on the three Bloch directions $(a,b,c)$. Moreover, $Q$ varies continuously with these directions: within each local region of parameter space, the eigenvectors can be chosen continuously; when different regions overlap, the resulting $Q$ agrees. Reversing any one Bloch direction interchanges the two amplitude products corresponding to even and odd parity. Therefore,
\begin{equation}
 Q(-a,b,c)=\overline{Q(a,b,c)}.
 \label{eq:oddQ}
\end{equation}

\begin{lemma}\label{lem:phase}
Suppose the chosen local bases satisfy
\begin{equation}
 \langle ZZI\rangle_w=\langle ZIZ\rangle_w=\langle IZZ\rangle_w=0.
 \label{eq:pairzero}
\end{equation}
Then there exists a local diagonal unitary transformation such that the state has an equal-modulus expansion in the GHZ basis if and only if $\im Q(w)=0$.
\end{lemma}
\begin{proof}
Normalization and Eq.~\eqref{eq:pairzero} give four linear equations. Inverting the corresponding $4\times4$ Walsh matrix, we obtain
\begin{equation}
 |w_x|^2+|w_{\bar x}|^2=\frac14,
 \qquad x\in\mathcal X.
 \label{eq:pairmass}
\end{equation}
Let $r_x=w_x\overline{w_{\bar x}}$, and denote them by $r_0,r_1,r_2,r_3$ in the order $x=000,001,010,011$. After applying the local diagonal unitary transformation
\begin{equation}
 D_\theta=\bigotimes_{j=1}^3
 \operatorname{diag}(e^{i\theta_j/2},e^{-i\theta_j/2}),
\end{equation}
$r_x$ becomes $e^{i s_x\cdot\theta}r_x$, where $s_x$ is the row vector corresponding to $x$ in the following matrix:
\begin{equation}
 T=\begin{pmatrix}1&1&1\\1&1&-1\\1&-1&1\\1&-1&-1\end{pmatrix}.
 \label{eq:signs}
\end{equation}
Denote these four row vectors by $s_0,s_1,s_2,s_3$ in order. Any three of them are linearly independent; the only independent linear relation among the four is $s_0-s_1-s_2+s_3=0$.

First consider the case in which all four $r_x$ are nonzero. For these four coherences to become purely imaginary after the transformation, the phases must satisfy
\begin{equation}
 s_x\cdot\theta=\frac\pi2-\arg r_x\pmod\pi.
 \label{eq:phaseeq}
\end{equation}
From the linear relation among the row vectors, this system of equations has a solution if and only if
\begin{equation}
 \arg r_0-\arg r_1-\arg r_2+\arg r_3=0\pmod\pi,
\end{equation}
that is, $\im(r_0r_3\overline{r_1r_2})=\im Q=0$. Specifically, it suffices to first solve any three equations in Eq.~\eqref{eq:phaseeq} over the reals; as long as the above compatibility condition holds, the fourth equation is then automatically satisfied modulo $\pi$.

If some $r_x$ is zero, no phase condition need be imposed on it. Then at most three equations remain, and by the linear independence of the corresponding row vectors, these equations always have a solution. Meanwhile, by definition, $Q=0$.

With the phases chosen in this way, the real parts of these four transformed coherences vanish. Combining this with Eq.~\eqref{eq:pairmass} gives
\begin{equation}
 |\langle G_{x,\pm}|D_\theta w\rangle|^2
 =\frac{|w_x|^2+|w_{\bar x}|^2}{2}=\frac18.
\end{equation}
Conversely, if the transformed state has an equal-modulus expansion in the GHZ basis, then $\re(e^{i s_x\cdot\theta}r_x)=0$ for every $x\in\mathcal X$. Since $Q$ is invariant under $D_\theta$, we must have $\im Q(w)=0$.
\end{proof}

Combining the above local basis transformation with the phase adjustment yields the local unitary transformation required in Theorem~\ref{thm:main}:
\begin{equation}
 V=D_\theta E^\dagger
   =D_\theta(E_A^\dagger\otimes E_B^\dagger\otimes E_C^\dagger).
 \label{eq:localunitary}
\end{equation}

\subsection{An identity for the correlation matrices}\label{sec:correlations}

In the following, we adopt a fixed Pauli coordinate system and compute all expectation values in the original state $\ket\psi$. Define three real correlation matrices
\begin{align}
 A_{ij}&=\langle\sigma_i\otimes\sigma_j\otimes I\rangle,\notag\\
 B_{ij}&=\langle\sigma_i\otimes I\otimes\sigma_j\rangle,\notag\\
 C_{ij}&=\langle I\otimes\sigma_i\otimes\sigma_j\rangle,
 \label{eq:correlations}
\end{align}
where $(\sigma_1,\sigma_2,\sigma_3)=(X,Y,Z)$. Then the three conditions in Eq.~\eqref{eq:pairzero} can be written as
\begin{equation}
 a^{\mathsf T}Ab=0,\qquad a^{\mathsf T}Bc=0,\qquad b^{\mathsf T}Cc=0.
 \label{eq:bilinear}
\end{equation}

\begin{lemma}\label{lem:symmetry}
For any three-qubit pure state, the matrix
\begin{equation}
 K=B^{\mathsf T}\adj(A)^{\mathsf T}C
 \label{eq:adj}
\end{equation}
is symmetric. If $A$ is invertible, let $M=A^{-1}B$; then
\begin{equation}
 D=M^{\mathsf T}C=B^{\mathsf T}A^{-\mathsf T}C
 \label{eq:D}
\end{equation}
is also symmetric.
\end{lemma}
\begin{proof}
Local unitary transformations correspond to rotations with determinant $1$ in Bloch space. Under these rotations, the three correlation matrices become
\begin{align}
 A&\mapsto R_AAR_B^{\mathsf T},&
 B&\mapsto R_ABR_C^{\mathsf T},&
 C&\mapsto R_BCR_C^{\mathsf T}.
\end{align}
The adjugate matrix satisfies
\begin{equation}
 \adj(R_AAR_B^{\mathsf T})^{\mathsf T}
 =R_A\adj(A)^{\mathsf T}R_B^{\mathsf T}.
 \label{eq:adj-covariance}
\end{equation}
When $A$ is invertible, this relation follows directly from the formula relating the adjugate matrix to the inverse matrix. Both sides are polynomials in the matrix entries; therefore, by continuity, the relation remains valid when $A$ is not invertible. Substituting it into the definition of $K$ gives $K\mapsto R_CKR_C^{\mathsf T}$. This shows that local unitary transformations preserve the symmetry of $K$.

Since any three-qubit pure state is locally unitary equivalent to a state in the Ac\'{i}n canonical form~\cite{Acin2001}, it suffices to consider
\begin{equation}
 \ket\psi=\ell\ket{000}+(u+iv)\ket{100}
              +p\ket{101}+q\ket{110}+r\ket{111},
 \label{eq:acin}
\end{equation}
where all parameters are real. Appendix~\ref{app:identity} lists the three correlation matrices corresponding to this form and gives explicit identities showing that the three pairs of off-diagonal entries of $K$ are equal. These polynomial identities do not require the normalization condition and also hold when parameters vanish; hence $K$ is symmetric. When $A$ is invertible, it follows from $D=K/\det A$ that $D$ is also symmetric.
\end{proof}

\subsection{Vanishing of a product amplitude}\label{sec:oddmap}

When $D$ is a positive definite or negative definite matrix, we shall use the following lemma. It shows that one can always choose suitable local directions such that one expansion coefficient in the corresponding product basis vanishes.

\begin{lemma}\label{lem:zero}
Let $L_1,L_2,L_3$ be invertible $3\times3$ real matrices. For $R\in\so(3)$, define three unit vectors
\begin{equation}
 n_i(R)=\frac{L_iRe_i}{\|L_iRe_i\|},\qquad i=1,2,3,
 \label{eq:deformed}
\end{equation}
where $e_1,e_2,e_3$ are the standard coordinate basis vectors. For any $\ket\psi\in(\C^2)^{\otimes3}$, there exists a rotation $R$ such that the tensor product of the $+1$ eigenstates of the three Pauli observables $n_i(R)\cdot\bm\sigma$ is orthogonal to $\ket\psi$.
\end{lemma}
\begin{proof}
We first construct, for $g\in\su(2)\simeq S^3$, continuous normalized spinors $v_i(g)$ whose Bloch vectors are $n_i(R(g))$ and which satisfy
\begin{equation}
 v_i(-g)=-v_i(g).
 \label{eq:oddspinor}
\end{equation}
Here $R(g)$ denotes the three-dimensional rotation corresponding to $g$.

Perform the polar decomposition $L_i=O_iP_i$ for $L_i$, where $P_i$ is positive definite and $O_i$ is orthogonal. Let $\delta_i=\det O_i\in\{1,-1\}$. Choose a fixed spinor $s_i$ whose Bloch vector is $\delta_i e_i$, and a fixed unitary matrix $U_i\in\su(2)$ corresponding to the rotation $\delta_i O_i$. The spinor $U_igs_i$ has Bloch vector $b_i=O_iR(g)e_i$. Under $g\mapsto-g$, the spinor changes sign, while its Bloch vector remains unchanged.

Let
\begin{equation}
 t_i=\frac{O_iP_iR(g)e_i}{\|P_iR(g)e_i\|}.
\end{equation}
Write $r_i=R(g)e_i$. Since $P_i$ is positive definite,
\begin{equation}
 b_i\cdot t_i=\frac{r_i^{\mathsf T}P_ir_i}{\|P_ir_i\|}>0.
 \label{eq:positive-overlap}
\end{equation}
Therefore, one can define the continuous matrix
\begin{equation}
 W_i(g)=\frac{I+(t_i\cdot\bm\sigma)(b_i\cdot\bm\sigma)}
 {\sqrt{2(1+b_i\cdot t_i)}}.
 \label{eq:transport}
\end{equation}
We now verify that $W_i(g)$ exactly transforms the spinor corresponding to the Bloch direction $b_i$ to the direction $t_i$. Write $B_i=b_i\cdot\bm\sigma$, $T_i=t_i\cdot\bm\sigma$, and let $N_i=I+T_iB_i$. Using Pauli matrix identities gives
\begin{align}
 N_i^\dagger N_i&=2(1+b_i\cdot t_i)I,\notag\\
 T_iN_i&=N_iB_i,\notag\\
 \det N_i&=2(1+b_i\cdot t_i).
 \label{eq:transport-identities}
\end{align}
It follows that $W_i$ is a unitary matrix with determinant $1$ and maps the $+1$ eigenspace of $B_i$ to the $+1$ eigenspace of $T_i$. Moreover, since $W_i(g)$ depends on $g$ only through $R(g)$, we have $W_i(-g)=W_i(g)$. Hence, setting $v_i(g)=W_i(g)U_igs_i$ yields continuous normalized spinors with the required Bloch vectors that satisfy Eq.~\eqref{eq:oddspinor}. The introduction of the sign $\delta_i$ makes this construction applicable both when $\det L_i$ is positive and when it is negative.

Now consider the continuous complex-valued function
\begin{equation}
 f(g)=\langle v_1(g)\otimes v_2(g)\otimes v_3(g)|\psi\rangle.
 \label{eq:oddfunction}
\end{equation}
When $g$ is replaced by $-g$, all three spinors change sign, so $f(-g)=-f(g)$. We now prove that $f$ must have a zero.

Suppose, for contradiction, that $f$ is nowhere zero. Since $S^3$ is simply connected, there exists a continuous real-valued argument function $h$ such that $f/|f|=e^{ih}$. From the oddness of $f$, we have
\begin{equation}
 h(-g)-h(g)\in\pi+2\pi\mathbb Z.
\end{equation}
The function $h(-g)-h(g)$ is continuous, and its values can only lie in the discrete set $\pi+2\pi\mathbb Z$. Since $S^3$ is connected, this function must be constant. But after replacing $g$ by $-g$, it changes sign; hence this constant can only be zero, which contradicts the fact that zero does not belong to $\pi+2\pi\mathbb Z$. Therefore, $f$ must have a zero, and the lemma is proved.
\end{proof}

The last step of the proof is a special case of the Borsuk--Ulam theorem. The explicit construction above guarantees that the spinors required for this argument can be chosen continuously.

\subsection{Invertible correlation matrices}\label{sec:cases}

First assume that $A,B,C$ are all invertible. Then $M=A^{-1}B$ is also invertible. By Lemma~\ref{lem:symmetry}, $D=M^{\mathsf T}C$ is a nonsingular real symmetric matrix.

If $D$ is indefinite, then there exists a unit vector $c$ such that $c^{\mathsf T}Dc=0$. Taking $b=Mc/\|Mc\|$, we obtain
\begin{equation}
 b^{\mathsf T}Cc=0,\qquad Ab=\frac{Bc}{\|Mc\|}.
\end{equation}
Fix these two directions $b,c$, and let the unit vector $a$ vary in the plane perpendicular to the nonzero vector $Bc$. Such $a$ form a unit circle, and every point on the circle satisfies the three conditions in Eq.~\eqref{eq:bilinear}. According to Eq.~\eqref{eq:oddQ}, the continuous function $\im Q$ takes opposite values at any two antipodal points $a$ and $-a$ on the circle. Therefore, by the intermediate value theorem, at least one point on the semicircular arc connecting these two points satisfies $\im Q=0$. Then, by Lemma~\ref{lem:phase}, one can choose a local diagonal unitary transformation such that Eq.~\eqref{eq:flat} holds.

If $D$ is positive definite or negative definite, let
\begin{equation}
 S=CM^{-1}=M^{-\mathsf T}DM^{-1}.
\end{equation}
Then $S$ is also a real symmetric matrix, and it is positive definite or negative definite together with $D$. Choose $\kappa\in\{1,-1\}$ such that $H=\kappa S>0$, and let $F=H^{1/2}$. For any right-handed orthonormal basis $(r_1,r_2,r_3)$, define
\begin{equation}
 \widetilde a=A^{-\mathsf T}Fr_1,\quad
 \widetilde b=F^{-1}r_2,\quad
 \widetilde c=M^{-1}F^{-1}r_3.
 \label{eq:frame}
\end{equation}
Since these linear maps are all invertible, the three vectors are all nonzero. A direct calculation gives
\begin{align}
 \widetilde a^{\mathsf T}A\widetilde b&=r_1^{\mathsf T}r_2=0,\notag\\
 \widetilde a^{\mathsf T}B\widetilde c&=r_1^{\mathsf T}r_3=0,\notag\\
 \widetilde b^{\mathsf T}C\widetilde c&=\kappa r_2^{\mathsf T}r_3=0.
 \label{eq:framezeros}
\end{align}
Normalizing them respectively, the resulting unit vectors $a,b,c$ still satisfy Eq.~\eqref{eq:bilinear}. Applying Lemma~\ref{lem:zero} to the three invertible linear maps in Eq.~\eqref{eq:frame}, one can choose a right-handed orthonormal basis such that the corresponding expansion coefficient $w_{000}=0$. In this case $Q=0$, and the three two-body correlation conditions also hold; hence Eq.~\eqref{eq:flat} follows from Lemma~\ref{lem:phase}. By choosing the sign of $\kappa$, this construction covers both the positive definite and negative definite cases of $D$.

\subsection{Singular correlation matrices}\label{sec:singular}

Finally, we deal with the case in which the correlation matrices are singular. First we show that pure states for which all three correlation matrices $A,B,C$ are invertible form a dense subset of all three-qubit pure states. For the normalized W state, we have
\begin{equation}
 A_W=B_W=C_W=\operatorname{diag}(2/3,2/3,-1/3),
\end{equation}
whose determinants are all $-4/27$. Given an arbitrary normalized vector $\psi$, perturb it to obtain $\psi+\epsilon W$. Before normalization, the correlation matrices computed from this vector have entries that are polynomials of degree at most two in the real parameter $\epsilon$. In the product of the three determinants, the coefficient of $\epsilon^{18}$ is $(-4/27)^3\ne0$, so this product is not the zero polynomial and can have only finitely many real roots. Therefore, as long as these roots are avoided, one can choose an arbitrarily small nonzero perturbation such that all three correlation matrices are simultaneously invertible. When $|\epsilon|<1/2$,
\begin{equation}
 \|\psi+\epsilon W\|\ge 1-|\epsilon|>1/2.
 \label{eq:perturbation-norm}
\end{equation}
This guarantees that the perturbed vector is nonzero and can be normalized. Normalization amounts to dividing each correlation matrix by the positive number $\|\psi+\epsilon W\|^2$, and hence does not destroy invertibility. The required density follows.

Now take a sequence of such normalized states $\psi_k\to\psi$, and for each $\psi_k$ choose local unitary operators $V_{A,k},V_{B,k},V_{C,k}$ satisfying Eq.~\eqref{eq:flat}. Since $U(2)^3$ is compact, one can select a common subsequence such that these three local unitary operators all converge. Here it is only necessary to choose one solution for each state; this choice is not required to depend continuously on the input state. The probabilities in Eq.~\eqref{eq:flat} depend continuously on the state and the local unitary operators; therefore, after taking the limit, all eight probabilities still equal $1/8$. This extends the conclusion to the case of singular correlation matrices and completes the proof of Theorem~\ref{thm:main} for all normalized three-qubit pure states.

\section{Applications}\label{sec:applications}

This section uses the orthonormal basis constructed in Eq.~\eqref{eq:encoding} to give two applications.

\subsection{Deterministic distribution over three Bell links}\label{sec:network}

Consider a central node $O$ and three spatially separated receivers $A,B,C$. The central node shares an ideal Bell pair with each receiver. Denote the three qubits held by the central node by $O_1,O_2,O_3$; then the total resource state is
\begin{align}
 \ket\Omega
 &=\ket{\Phi^+}_{O_1A}\ket{\Phi^+}_{O_2B}\ket{\Phi^+}_{O_3C}\notag\\
 &=\frac1{\sqrt8}\sum_{x\in\{0,1\}^3}\ket{x}_O\ket{x}_{ABC},
 \label{eq:resource}
\end{align}
where $\ket{\Phi^+}=(\ket{00}+\ket{11})/\sqrt2$.

Given a target state $\psi$, choose the corresponding encoding operations $U_s$ according to Eq.~\eqref{eq:encoding}. Since the encoded states $\psi_s$ form an orthonormal basis, their complex conjugates in the computational basis also form an orthonormal basis. Therefore, the central node can perform the following rank-one projective measurement on its three qubits:
\begin{equation}
 \Pi_s=\ket{\psi_s^*}\bra{\psi_s^*}.
 \label{eq:measurement}
\end{equation}
When the measurement outcome is $s$, the unnormalized conditional state of the receivers is
\begin{equation}
 (\bra{\psi_s^*}_O\otimes I_{ABC})\ket\Omega
 =\frac1{\sqrt8}\ket{\psi_s}_{ABC}
 =\frac1{\sqrt8}U_s\ket\psi.
 \label{eq:networkbranch}
\end{equation}
Each outcome occurs with probability $1/8$. After the central node broadcasts the three-bit label $s$ to the receivers, each party applies the local factor belonging to them in $U_s^\dagger$, and together they obtain the target state $\psi$. This correction is valid for all measurement outcomes, so the protocol is deterministic, and the receivers only need to perform local operations.

This is an entanglement-swapping protocol for any prescribed three-qubit pure state. Theorem~\ref{thm:main} guarantees the existence of the required projective measurement and commuting-reflection corrections. For each outcome, the state before correction is locally unitarily equivalent to the target, so the two states have the same entanglement spectrum across every bipartition.

The target state and correction table are agreed in advance. The three-bit message specifies only the measurement outcome. Three Bell pairs are sufficient resources for completing this protocol, but for some specific target states fewer resources may be needed.

\subsection{Minimum label entropy for randomizing a known pure state}\label{sec:randomization}

We measure the randomness cost by the Shannon entropy of the full classical label required to select the overall unitary operation. Private random choices made by the parties must also be included in this label.

Suppose the three parties jointly hold a known pure state $\psi$ and share a uniformly distributed three-bit random variable $s$. According to $s$, each party applies the corresponding local factor in $U_s$. After discarding the random label, it follows from Eq.~\eqref{eq:gram} that the average state of the three parties is
\begin{equation}
 \frac18\sum_s U_s\ket\psi\bra\psi U_s^\dagger
 =\frac{I_8}{8}.
 \label{eq:randomization}
\end{equation}
Thus product reflections and three bits of shared randomness suffice to map the known pure state to the maximally mixed state.

We now show that no further reduction in the required label entropy is possible. Consider an arbitrary set of unitary operators $W_j$ and probabilities $p_j$ satisfying
\begin{equation}
 \sum_jp_jW_j\ket\psi\bra\psi W_j^\dagger=\frac{I_8}{8}.
 \label{eq:anyrandomization}
\end{equation}
The output of each branch is a pure state, and the entropy of the mixed state of a pure-state ensemble does not exceed the Shannon entropy of its probability distribution. Therefore, taking logarithms base $2$, we have
\begin{equation}
 3=S(I_8/8)\le H(p).
 \label{eq:entropybound}
\end{equation}
At the same time, the rank of the mixed state does not exceed the number of branches with nonzero probability, so at least eight such branches are required. The scheme in Eq.~\eqref{eq:randomization} uses exactly eight equally probable branches and simultaneously attains the lower bounds on both label entropy and number of branches. Since the derivation of the above entropy lower bound does not use any locality condition, local unitary operations can already achieve the minimum label entropy attainable by arbitrary global unitary operations.

The bound applies to the entropy of the full label, including all private random choices. The availability of free private randomness does not, by itself, imply that three bits of shared randomness are necessary; the unitary table may also depend on the known input state. This differs from universal quantum encryption, which requires one fixed channel to randomize every input state and has a different key cost~\cite{Ambainis2000}.

\section{Discussion}\label{sec:discussion}

We prove that any three-qubit pure state, after suitable local unitary transformations, can be expanded with equal moduli in the same fixed GHZ basis. Combined with the four-qubit counterexample in Ref.~\cite{Pauwels2026}, this shows that complete local encoding is achievable for all three-qubit pure states, whereas in four-qubit systems there already exist cases in which it cannot be achieved.

The resulting basis gives locally correctable state distribution and a randomization protocol with minimum label entropy.

Our result is an existence proof; we do not yet provide an efficient, stable construction that works for all input states. Two issues remain open for further investigation: how to determine the required local unitary transformation in a stable manner when the correlation matrices are nearly singular, and how many resources are needed to implement the joint measurement.

\onecolumngrid
\clearpage
\appendix
\section{Algebraic derivation of the correlation identity}\label{app:identity}

Without normalizing the vector in Eq.~\eqref{eq:acin}, we directly compute the corresponding Pauli correlations and obtain
\begin{align}
 A&=\begin{pmatrix}
 2q\ell&0&2u\ell\\
 0&-2q\ell&2v\ell\\
 -2uq-2pr&2vq&-u^2-v^2-p^2+q^2+r^2+\ell^2
 \end{pmatrix},\label{eq:Aexplicit}\\
 B&=\begin{pmatrix}
 2p\ell&0&2u\ell\\
 0&-2p\ell&2v\ell\\
 -2up-2qr&2vp&-u^2-v^2+p^2-q^2+r^2+\ell^2
 \end{pmatrix},\label{eq:Bexplicit}\\
 C&=\begin{pmatrix}
 2ur+2pq&-2vr&2uq-2pr\\
 -2vr&-2ur+2pq&-2vq\\
 2up-2qr&-2vp&u^2+v^2-p^2-q^2+r^2+\ell^2
 \end{pmatrix}.\label{eq:Cexplicit}
\end{align}
To simplify the expressions, define
\begin{align}
 \nu&=\ell^2+u^2+v^2+p^2+q^2+r^2,\notag\\
 h&=q(\nu-2p^2)+2upr,\notag\\
 j&=upq(\nu-2p^2-2r^2)+q^2r(u^2+v^2-p^2+q^2-r^2-\ell^2)-2v^2p^2r.
 \label{eq:polynomials}
\end{align}
Substituting these matrices into the definition of $K$ in Eq.~\eqref{eq:adj} gives
\begin{align}
 K_{12}=K_{21}&=8vpr\ell^2h,\notag\\
 K_{23}=K_{32}&=8vpq\ell^2h,\notag\\
 K_{13}=K_{31}&=-8q\ell^2j.
 \label{eq:offdiag}
\end{align}
The above identities hold for arbitrary values of the six real parameters. No division is used in the derivation, so no separate discussion is needed when the parameters vanish. Since the three pairs of off-diagonal entries are respectively equal, $K$ is a symmetric matrix. A direct calculation also gives $\det A=-4q\ell^2h$, which is consistent with the relation $D=K/\det A$ when $A$ is invertible.

\twocolumngrid

\end{document}